\documentstyle[12pt,epsfig]{article}
\newcommand{\singlespace}{
    \renewcommand{\baselinestretch}{1}\large\normalsize}

\newcommand{\be}{\begin{equation}}
\newcommand{\ee}{\end{equation}}
\newcommand{\ba}{\begin{eqnarray}}
\newcommand{\ea}{\end{eqnarray}}

\newcommand{\ave}[1]{\langle {#1} \rangle}

\def\qb{\bar q}

\def\roughly#1{\mathrel{\raise.3ex\hbox{$#1$\kern-.75em%
\lower1ex\hbox{$\sim$}}}}
\def\lsim{\roughly<}

\begin{document}
%
\vspace{1.0in}
\begin{flushright}
 April 1999
\end{flushright}
\vspace{1.0in}
\begin{center}
\singlespace
\begin{large}
{\bf Strange quark matter with dynamically generated quark masses}\\
\end{large}
\vskip 1.0in
M. Buballa\footnote{E-mail: michael.buballa@physik.tu-darmstadt.de}
and M. Oertel\footnote{E-mail: micaela.oertel@physik.tu-darmstadt.de}\\
{\small{\it Inst. f. Kernphysik, TU Darmstadt,\\ 
Schlossgartenstr. 9, 64289 Darmstadt, Germany}}\\
\end{center}
\vspace{2cm}
\begin{abstract}
 Bulk properties of strange quark matter (SQM) are investigated within the
 SU(3) Nambu--Jona-Lasinio model. In the chiral limit the model behaves
 very similarly to the MIT bag model which is often used to describe SQM.
 However, when we introduce realistic current quark masses, the strange 
 quark becomes strongly disfavored, because of its large dynamical mass. 
 We conclude that SQM is not absolutely stable.


\end{abstract}
\newpage

\singlespace

The properties of strange quark matter (SQM) and in particular the conjecture
that SQM could be the absolute ground state of strongly interacting matter
\cite{Bodmer,Witten} have attracted much attention in nuclear physics and 
astrophysics. 
Whereas the empirical stability of ordinary nuclei excludes the existence of
absolutely stable non-strange quark matter (NSQM) this argument does not
hold for SQM because the decay of nuclei into SQM would involve higher-order
weak interaction processes, associated with a very long lifetime.
In 1984 Farhi and Jaffe \cite{Farhi} investigated the question of absolutely 
stable SQM within the MIT bag model \cite{MIT}. Treating the bag constant
and the strange quark mass as free parameters they found a reasonable
window for which SQM is stable and NSQM is unstable compared with an 
$^{56}Fe$-nucleus. Almost 15 years later  
bag model calculations still play a key role in this field (for reviews 
see \cite{SQM1}-\cite{SQM4}). 

In ref.~\cite{Buballa} we have investigated non-strange 
quark matter within the Nambu--Jona-Lasinio (NJL) model \cite{NJL} in mean
field approximation. For vanishing (current) quark masses and with the
appropriate choice of parameters we found that quark matter in the 
NJL mean field behaves very similarly to quark matter in an MIT bag. For
the description of SQM we have to introduce finite quark masses,
at least in the strange quark sector. In this case the correspondence 
between the two models becomes less accurate, with the NJL mean field
behaving in a much more complex way. This is related to the fact that
the bag constant in the NJL model is not a phenomenological input parameter,
like in a bag model, but a dynamical property, which has its origin in
the spontaneous breaking of chiral symmetry.  
This motivated us to extend the model of ref.~\cite{Buballa}
to investigate bulk properties of strange quark matter within the NJL model. 

The three-flavor NJL model has been discussed by many authors, e.g. 
\cite{HK}-\cite{Rehberg}. In this article we adopt the Lagrangian of 
ref.~\cite{Rehberg}:
\ba
{\cal L} \;=\; \qb ( i \partial{\hskip-2.0mm}/ - {\hat m_0}) q
            \;&+&\; G \sum_{k=0}^8 [\,(\qb\lambda_k q)^2 + 
           (\qb i\gamma_5\lambda_k q)^2\,] \nonumber \\
            \;&-&\; K \,[ \,det_f (\qb(1+\gamma_5) q) 
                         + det_f (\qb(1-\gamma_5) q) \,]   \ .
\label{L3}
\ea
Here $q$ denotes a quark field with three flavors, $u$, $d$ and $s$, and
three colors. ${\hat m_0} = diag(m_0^u,m_0^d,m_0^s)$ is a $3 \times 3$ matrix 
in flavor space. We restrict ourselves to the isospin-symmetric case, 
$m_0^u=m_0^d$. The Lagrangian contains a $U(3)_L \times U(3)_R$-symmetric
four-point interaction term and a six-point interaction, which is a determinant
in flavor space and which breaks the $U_A(1)$-symmetry. $G$ and $K$ are 
constants with dimension energy$^{-2}$ and energy$^{-5}$, respectively.

The model is not renormalizeable and we have to specify a regularization 
scheme for divergent integrals. For simplicity we use a sharp cut-off
$\Lambda$ in 3-momentum space.  Besides $\Lambda$ we have to fix
four parameters, namely the coupling constants $G$ and $K$ and
the current quark masses $m_0^u$ and $m_0^s$. In most of our calculations
we will adopt the parameters of ref.~\cite{Rehberg}: $\Lambda$ = 602.3 MeV,
$G\Lambda^2$ = 1.835, $K\Lambda^5$ = 12.36, $m_0^u$ = 5.5~MeV and
$m_0^s$ = 140.7~MeV. These parameters have been determined by fitting
$f_\pi$, $m_\pi$, $m_K$ and $m_{\eta'}$ to their empirical values,
while the mass of the $\eta$-meson is  underestimated by about 6\%. 

Due to the interactions the quarks acquire dynamical masses $m_i$ 
which are in general different from their current masses $m_0^i$ and 
depend on temperature and density. In the following we consider quark
matter at $T = 0$ and zero or non-zero baryon number density 
$\rho_B = \frac{1}{3} 
n_B = \frac{1}{3} (n_u + n_d + n_s)$, with $n_i = \ave{q_i^\dagger q_i}$. 
Then, in mean field (Hartree) 
approximation, the dynamical quark masses  are solutions of the gap equation
\be
   m_i \;=\; m_0^i \;-\; 4 G \,\ave{\bar{q}_i q_i} \;+\;
           2 K \, \ave{\bar{q}_j q_j} \ave{\bar{q}_k q_k}  \ ,
\label{gap}
\ee   
with $(q_i,q_j,q_k)$ being any permutation of $(u,d,s)$. The quark 
condensates $\ave{\bar{q}_i q_i}$ are given by
\be
   \ave{\bar{q}_i q_i} \;=\; - \frac{3}{\pi^2} \int_{{p_F}_i}^\Lambda
   p^2 dp  \; \frac{m_i}{\sqrt{m_i^2 + p^2}} \ . 
\label{cond}
\ee   
They depend on the Fermi momenta ${p_F}_i = (\pi^2 n_i)^{1/3}$ and again
on the dynamical masses $m_i$. Thus, for given quark number densities $n_i$, 
eqs.~(\ref{gap}) and (\ref{cond}) have to be solved self-consistently. 
After that we can calculate the energy density $\varepsilon$ and the total 
pressure $p$ of the system:
\be
   \varepsilon \;=\; \sum_{i=u,d,s} \; \frac{3}{\pi^2}
    \int_0^{{p_F}_i} p^2 dp \; \sqrt{m_i^2 + p^2} 
   \;-\; (B - B_0) \ , \qquad 
   p = - \varepsilon \;+ \sum_{i=u,d,s} n_i \,\sqrt{m_i^2 + {p_F}_i^2} \ .
\label{epsilon}
\ee
Here $B$ denotes the bag pressure,
\be
   B \;=\; \sum_{i=u,d,s} (\; \frac{3}{\pi^2}
    \int_0^\Lambda p^2 dp \; (\sqrt{m_i^2 + p^2} \;-\; 
   \sqrt{{m_0^i}^2 + p^2}) \;-\; 2 G  \ave{\bar{q}_i q_i}^2 \;)
   \;+\; 4 K \, \ave{\bar u u} \ave{\bar d d} \ave{\bar s s}
   \ ,
\label{B}
\ee
and $B_0 = B|_{n_u=n_d=n_s=0}$. The latter was introduced in
eq.~(\ref{epsilon}) in order to ensure $\varepsilon = p = 0$ in vacuum.   

The existence of a bag pressure term in eq.~(\ref{epsilon}) suggests
that there is a certain connection between the present model and 
standard bag model descriptions of quark matter. Let us have a closer
look at this point. It is useful to begin with the chiral limit, 
i.e. $m_0^i$ = 0 for all flavors.
With $\Lambda$, $G$ and $K$ as specified above, chiral symmetry is 
spontaneously broken in vacuum with equal dynamical masses
$m_u = m_d = m_s$ = 310.6 MeV and a bag pressure $B_0 = 57.3 \rm{MeV / fm^3}$. 
(There are other solutions of the gap equation but they are energetically 
less favored.) 
Now we consider $SU(3)$-symmetric quark matter, $n_u = n_d = n_s = \rho_B$.
At low densities the dynamical quark masses decrease with density but
they remain finite. However, when $\rho_B$ exceeds a critical value
of  $\sim 0.9 \rho_0$ chiral symmetry gets restored and the quark masses
and condensates vanish. Consequently the bag pressure is also zero in 
this regime and we get from eq.~(\ref{epsilon}) 
\be
   \varepsilon \;=\; \frac{9}{4\pi^2} {p_F}^4 \;+\; B_0
   \ , \qquad p \;=\; \frac{3}{4\pi^2} {p_F}^4 \;-\; B_0 \;.
\label{eps3}
\ee
This means the system behaves exactly like a gas of massless non-interacting 
quarks with three flavors and three colors inside a large MIT bag with a 
bag constant $B_0$. 
It should be kept in mind, however, that eq.~(\ref{eps3}) only holds in the
regime of vanishing dynamical quark masses. For the present parameters it
turns out that the minimal energy per baryon number $E/A = \varepsilon/\rho_B$
is indeed found inside this region. Thus eq.~(\ref{eps3}) is appropriate
to describe the system in the vicinity of the point of stability.

For comparison we consider  non-strange isospin-symmetric quark matter,
$n_s=0$, $n_u = n_d = \frac{3}{2} \rho_B$, still in the chiral limit. 
In this case strange and non-strange quarks behave differently with
density, and chiral symmetry is only partially restored at high densities:
When $\rho_B$ exceeds $\sim 1.0 \rho_0$ the non-strange quarks become
again massless but the strange quark remains massive with a mass
$m_s^*$ which is 154.2 MeV in our example. Hence the bag pressure remains 
also finite, $B = B^* = 3.7 \rm{MeV / fm^3}$, and the system now behaves like 
a bag model 
for a gas of massless quarks with two flavors and three colors but with an
effective bag constant $B_{eff} = B_0 - B^*$, which is smaller than the bag 
constant $B_0$ in the SU(3)-symmetric case. 

Thus, in contrast to a naive bag model, the effective bag constant which
describes the behavior of NJL quark matter depends on the flavor composition
of the matter. This is also shown in fig.~1 where various properties
of isospin-symmetric quark matter are plotted as functions of the
fraction of strange quarks, $r_s = n_s / n_B$. 
With the present parameters stable quark matter exists in the region
$0 < r_s < 0.48$. As discussed in ref.~\cite{Buballa} these solutions
correspond to the high-density phase of a first-order chiral phase transition
in equilibrium with the non-trivial vacuum.
Throughout this paper we will denote the dynamical quark masses of these stable
matter solutions by $m_i^*$ and the corresponding bag pressure by $B^*$.
In fig.~1(a) the solid line indicates the energy per baryon number, $E/A$ of
stable quark matter. $m_i^*$ and the effective bag constant
$B_{eff} = B_0 - B^*$ are shown in fig.~1(b) and (c), respectively. 
For $0.12 < r_s < 0.48$ the system behaves like a bag model with massless 
quarks and a bag constant $B_{eff} = B_0$. 
For $r_s < 0.12$ the effective bag constant is somewhat smaller and the 
strange quarks become massive. 
In panel (a) we also show the bag model result for $E/A$ with a bag
constant $B_{BM}=B_0$ and massless quarks (dashed line). For
$0.12 < r_s < 0.48$ it agrees of course exactly with the NJL result, but
also for $r_s < 0.12$ the deviations are relatively small.
In fig.~1 all NJL curves stop at $r_s \simeq 0.48$.
For $r_s > 0.48$ the NJL quark matter is unstable against evaporation of
free massive s-quarks. In fact, this kind of phase separation is very 
typical for asymmetric matter and has been discussed for nuclear matter 
\cite{Barranco,MS} as well as for models of the deconfinement phase transition 
\cite{Glendenning,Mueller}. However, the evaporation of free quarks is of
course unphysical and reflects the missing confinement of the NJL model.     

With finite current quark masses $m_0^i$ things change considerably.
Fig.~2 shows the same quantities as fig.~1, but now
with $m_0^u = m_0^d$ = 5.5 MeV and $m_0^s$ = 140.7 MeV \cite{Rehberg}.
We now find stable quark matter for $0 < r_s < 0.79$.
In vacuum the dynamical masses of the up- and down-quarks are increased
by $\sim$~60~MeV to 367.6~MeV and that of the strange quark by
$\sim$~240~MeV to 549.5~MeV. 
Perhaps more important for us is the well-known fact that the quark
condensates do not completely vanish even at higher densities. As a
consequence, the dynamical quark masses in stable quark matter, $m_i^*$, are 
still well above the current masses $m_0^i$ (see fig.~2(b)). 
For non-strange isospin-symmetric quark matter ($r_s = 0$) we find 
$m_u^*$ = 52.6~MeV and $m_s^*$ = 464.4~MeV. Since on the other hand the
Fermi energy in this system is only 361.1~MeV the conversion of a non-strange
quark into a strange quark is obviously energetically not favored. 
Here the notion of dynamical quark masses is crucial: In a naive bag model
one would assign the strange quark its current mass which in our example
is more than 200 MeV lower than the Fermi energy. 

The difference becomes
obvious in fig.~2(a) where the energy per baryon number of stable
quark matter is plotted as a function of $r_s$. 
The solid line indicates the NJL result while the dotted line corresponds
to a bag model calculation with $B_{BM}$ = 105.2 $\rm{MeV/ fm^3}$ (the 
effective bag constant at $r_s = 0$) and quarks with $m_u = m_d$ = 5.5 MeV and 
$m_s$ = 140.7 MeV, our current quark masses. Whereas the latter has a minimum 
at a finite fraction of strange quarks, $r_s = 0.27$, the former is a strictly
rising function of $r_s$: additional strangeness is always disfavored in
the NJL calculation. In fig.~2(a) we also show the result of a
bag model calculation with $m_u = m_d$ = 52.6~MeV and $m_s$ = 464.4~MeV,
our dynamical quark masses at $r_s = 0$ (dashed-dotted line). This curve rises
even steeper with $r_s$ than the NJL result. The reason is that in the NJL
calculation the dynamical strange quark mass $m_s^*$  drops with $r_s$
(see fig.~2(b)). Therefore the penalty 
for adding strangeness becomes smaller than it would be with a constant mass 
$m_s^*(r_s=0)$, even though this effect is partially compensated by the fact 
that the effective bag constant $B_{eff} = B_0-B^*$ rises with $r_s$ for not 
too large $r_s$.   

$B_{eff}$ is shown in fig.~2(c). It varies much stronger with $r_s$ 
than in the chiral limit and is much larger in magnitude. Already at its 
minimum, at $r_s=0$, it is about twice as large. To major extent this is the 
reason why the energy per baryon number at this point is almost 200 MeV 
larger than in the chiral limit. Both, the magnitude of the effective bag 
constant and its strong variation with $r_s$, have their origin in 
the fact that the bag pressure is very sensitive to the dynamical quark 
masses. As a consequence the vacuum bag pressure $B_0$ is now 
291.7~${\rm MeV / fm^3}$, five times the chiral limit value, and, since
the strange quark mass in stable matter varies, the corresponding bag pressure 
$B^*$ varies strongly with $r_s$.

It should be also noted that, even for fixed $r_s$, quark masses and the bag 
pressure are density dependent and the values for $m_i^*$ and $B_{eff}$
shown in fig.~2 only correspond to the densities of stable quark 
matter. As soon as one moves away from the minimum of $E/A$, $m_i$ and
$B_0-B$ change. 
Thus, whereas in the chiral limit NJL quark matter could be well described 
in terms of an MIT-like bag model with vanishing quark masses  and a single 
bag constant (at least within a larger range of densities and strangeness 
fractions), it behaves in a much more complex way when we include finite 
current quark masses.

So far, focusing on the bag model aspect, we made some simplifying 
assumptions which we have to abandon for a more realistic study of SQM.
In particular we have to take into account weak decays. This implies that
we have to include electrons and (in principle) neutrinos.
To large extent we will adopt the model of Farhi and Jaffe \cite{Farhi}, 
with the MIT bag replaced by the NJL mean field. 
With the usual assumption that the neutrinos can freely leave the system,
the matter is characterized by four (three quark- and one electron-)
chemical potentials. They are related to the corresponding densities
in the standard way:
\be
   n_i \;=\;  \frac{1}{\pi^2}\,(\mu_i^2 - m_i^2)^{3/2} \,
               \theta(\mu_i^2 - m_i^2)  \quad {\rm for}  \quad i = u,d,s  
   \qquad {\rm and} \qquad n_e \;=\;  \frac{\mu_e^3}{3\pi^2}  \ .
\label{chemdens}
\ee
Here we neglected the electron mass. For the quarks one has to keep 
in mind that the masses entering the r.h.s. have to be calculated from the 
gap equation and are therefore density dependent themselves.
Treating the electrons as a free gas, energy density and pressure of
SQM are given by
\be
   \varepsilon_{SQM} \;=\; \varepsilon \;+\;\frac{\mu_e^4}{4\pi^2}
   \ , \qquad p_{SQM} \;=\; p \;+\;\frac{\mu_e^4}{12\pi^2} \ ,
\label{epssqm} 
\ee
with $\varepsilon$ and $p$ as defined in eq.~(\ref{epsilon}). 

In chemical equilibrium maintained by weak interactions only two of the four
chemical potentials are independent:
\be
   \mu_d \;=\; \mu_s \;=\; \mu_u \,+\, \mu_e  \;.
\label{chem}
\ee  
Furthermore we demand charge neutrality,
\be
   \frac{2}{3}\,n_u \;-\; \frac{1}{3}\,(n_d \,+\, n_s) \;-\; n_e \;=\; 0 \ .
\label{charge}
\ee  
Thus the system can be characterized by one independent quantity, e.g.
the baryon number density $\rho_B$.  

Our results are shown in fig.~3. Panel (a) shows the energy per
baryon number $E/A = \varepsilon_{SQM}/\rho_B$ as a function of $\rho_B$. 
The solid line corresponds to the model described above, the dotted line
to non-strange quark matter, where $\mu_s$ was set equal to zero by hand.
Obviously the strangeness degree of freedom is only important at densities
above $\sim~4\,\rho_0$. This can also be seen in panel (c) where the
fractions $r_i = n_i/n_b$ of the various particles are plotted. Since
electrons (dotted line) play practically no role (and are hardly visible
in the plot) the fraction of u-quarks (dashed-dotted) is fixed by charge
neutrality (eq.~(\ref{charge})) to $r_u \simeq$ 1/3. For
$\rho < 3.85\,\rho_0$ the fraction of strange quarks, $r_s$, (solid)
is zero and thus $r_d$ must be the remaining 2/3 (dashed). For 
$\rho > 3.85\,\rho_0$ $r_s$ becomes non-zero and $r_d$ drops accordingly.    

The fact that there is no strangeness at lower densities is again due to
the relatively large mass of the strange quark. The dynamical quark masses
are plotted in fig.~3(b) as functions of $\rho_B$. For comparison
we also show the chemical potential $\mu_s=\mu_d$ (dotted line). As long
as $\mu_s \leq m_s$ the density of strange quarks is zero.
For $\rho_B \lsim 2 \rho_0$, $m_s$ drops with density. This can be mainly
attributed to the decrease of $\ave{\bar u u}\ave{\bar d d}$ due to 
the rising density of non-strange quarks. At higher densities  
$\ave{\bar u u}\ave{\bar d d}$ is practically zero and $m_s$ stays almost
constant until $n_s$ becomes non-zero and causes $|\ave{\bar s s}|$ to
drop. This behavior is rather different from standard parameterizations
which have been used in the literature to study SQM and which depend on
the total baryon number density $\rho_B$ only \cite{Chak,Benv}.

At those densities where strange quarks exist, they lead to a reduction of
the energy, as can be seen by comparing the solid curve in fig.~3(a) 
with the dotted one. However, the minimum of $E/A$ does not lie in this 
regime, but at a much lower density, $\rho_B = 2.25\,\rho_0$.
Here we find $E/A$ = 1102 MeV. Compared with the energy per baryon
in an iron nucleus, $E/A \simeq$ 930 MeV, this is still very large. 
In this sense our results are consistent with the empirical fact that
stable NSQM does not exist. 
However, since the energy of strange quark matter is even higher, our
calculation predicts that also SQM is not the absolute ground state of
strongly interacting matter. 

This result is very robust with respect to changes of the model parameters.
Obviously stable quark matter with finite strangeness is only possible
if the in-medium strange quark mass $m_s^*$ is much lower than the value we 
obtained above. The easiest way to achieve this is to choose a lower 
value of the current mass $m_0^s$. If we leave all other parameters unchanged, 
$m_0^s$ should be at most 85 MeV if we require $n_s \neq 0$ at the 
minimum of $E/A$. With this value we obtain much too low masses for 
$K$ and $\eta$ ($m_K \simeq$ 390 MeV, $m_\eta \simeq$ 420~MeV), while
$E/A$ is still relatively large (1075 MeV). If we want to come down to
$E/A \simeq$ 930 MeV we have to choose $m_0^s$ = 10 MeV or, alternatively,
$m_0^s$ = 25 MeV and $m_0^u = m_0^d$ = 0. This is of course completely
out of range.

We could also try to lower $m_s^*$ by choosing a smaller coupling 
constant $G$. (Since $\ave{\bar u u}\ave{\bar d d}$ is already very small at 
the densities of interest, $m_s^*$ is almost insensitive to $K$). However,
with a lower $G$ the vacuum masses of all quarks drop and correspondingly
the vacuum bag pressure $B_0$. This causes the minimum of $E/A$ to move
to lower densities. It turns out that the corresponding chemical potential 
$\mu_s^*$ drops even faster than $m_s^*$ and it is thus not possible to 
obtain stable SQM in that way. 
To avoid the strong density decrease we could increase the coupling constant 
$K$ while decreasing $G$, e.g. in such a way that the vacuum masses of
u- and d-quarks are kept constant. In order to get $n_s \neq 0$ at the 
minimum of $E/A$ we have to lower $G\Lambda^2$ to 1.5 and to increase
$K\Lambda^5$ to 21.27, almost twice the value of our original parameter set.
For these parameters the energy per baryon number is still 1077 MeV. On
the other hand we cannot further decrease the ratio $G/K$ because that would
flip the sign of the effective $q\bar q$ coupling in the pseudoscalar-flavor
singlet channel, which is dominated by the combination
$2G + \frac{2}{3} K (\ave{\bar u u}+\ave{\bar d d}+\ave{\bar s s})$. 
In that case there would be no solution for the $\eta'$-meson in vacuum.
Hence there seems to be no realistic way to find absolutely stable SQM within 
our model. 

We could ask whether this result can change if we introduce
additional interaction terms to the Lagrangian eq.~(\ref{L3}). 
In particular vector interactions are known to be important in dense matter.
However, since vector mean fields
are repulsive, the energy per baryon number will be even larger than before
and SQM remains strongly disfavored compared with ordinary nuclei.
The same is true for the medium effects described in ref.~\cite{Schertler} 
which also tend to increase the energy. 
In fact, these effects are complementary to ours because they are most 
important (and most reliable) at high densities, whereas our dynamical
quark masses are most important at low densities.
In ref.~\cite{Jaminon} the chiral phase transition was studied within a 
``scaled'' NJL model. There is no six-point interaction in this model,
but the different flavors are coupled through a dilaton field. The
coupling was found to be weak except for rather low choices of the
gluon condensate \cite{VdB}. This makes it very unlikely to find absolutely
stable SQM within that model, although a quantitative study of this question
has not yet been done. 
  
In summary, we investigated strange and non-strange quark matter within the
NJL model. In the chiral limit the model behaves very similarly to the
MIT bag model which has been mostly used in the literature to describe SQM.
For realistic model parameters including finite current quark masses the
NJL model shows a more complex behavior.
This is due to the fact that bag pressure and effective
quark masses are dynamical properties of the model and are therefore in
general density dependent.  
We find that the dynamical mass
of the strange quark stays very large at the relevant densities. As a
consequence SQM is not favored compared with NSQM in the model and there
seems to be no chance to find SQM with energies per baryon number lower
than in ordinary nuclei. This almost rules out the original idea
of absolutely stable SQM as a simple consequence of the Pauli exclusion
principle. Of course we cannot completely exclude that more sophisticated
mechanisms, like color superconductivity \cite{ARW,Rapp}, in particular the 
recently discovered mechanism of color-flavor locking \cite{Alford},  
change our results. This is, however, beyond the scope of this letter.  
\pagebreak
\begin{center}
{\large Acknowledgement:}
\end{center}

We would like to thank J. Schaffner-Bielich for having drawn our attention 
to the topic of strange quark matter.

\vspace{2cm}


\phantom{empty}
\pagebreak

\pagestyle{empty}
\begin{figure}[h!]
\begin{center}
\parbox{9cm}{\epsfig{file=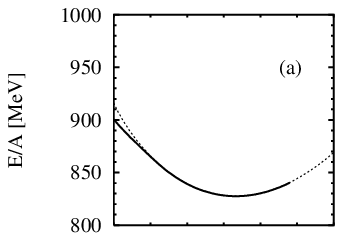, height=6.5cm, width=8.45cm}}
\vskip-12.55mm\hskip8.2mm
\parbox{9cm}{\epsfig{file=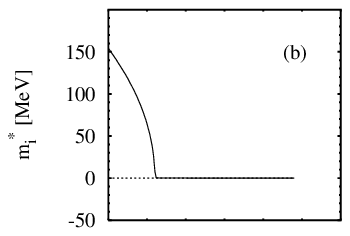, height=6.5cm, width=8.01cm}}
\vskip-13.30mm\hskip15.5mm
\parbox{9cm}{\epsfig{file=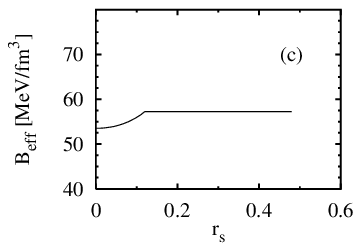, height=7.5cm, width=7.60cm}}
\caption{Properties of isospin-symmetric ($n_u=n_d$) stable quark matter 
      as a function of $r_s = n_s / n_B$ 
      in the chiral limit (parameters of ref.~\cite{Rehberg}, but with
      $m_0^i$ = 0).
      (a)~Energy per baryon number. The result of the present model
      (solid line) is compared with a bag model calculation with bag constant
      $B_{BM} = 57.3 \,MeV/ fm^3$ (dotted). 
      (b)~Dynamical masses of the strange (solid) and non-strange quarks 
      (dotted). 
      (c)~Effective bag constant.
}
\end{center}
\label{fig1}
\end{figure}

\begin{figure}[h!]
\begin{center}
\parbox{9cm}{\epsfig{file=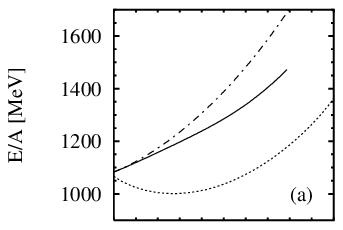, height=6.5cm, width=8.45cm}}
\vskip-12.55mm\hskip8.3mm
\parbox{9cm}{\epsfig{file=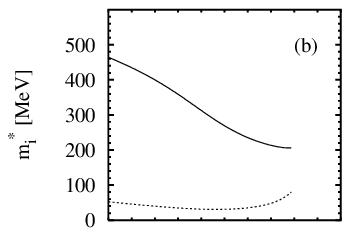, height=6.5cm, width=8.00cm}}
\vskip-13.30mm\hskip8.4mm
\parbox{9cm}{\epsfig{file=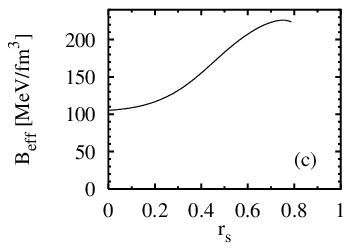, height=7.5cm, width=8.01cm}}\\
\caption{The same as fig.~1 but with finite current masses
      (parameters of ref.~\cite{Rehberg}). In panel (a) the
      result of the present model (solid line) is compared with bag model
      calculations with bag constant $B_{BM} = 105.2\,MeV/ fm^3$ and 
      $m_u = m_d$ = 5.5 MeV, $m_s$ = 140.7 MeV (dotted) or 
      $m_u = m_d$ = 52.6 MeV, $m_s$ = 464.4 MeV (dashed-dotted). 
}
\end{center}
\label{fig2}
\end{figure}

\begin{figure}[h!]
\begin{center}
\parbox{9cm}{\epsfig{file=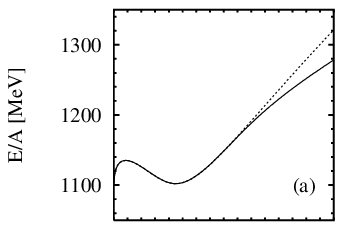, height=6.5cm, width=8.45cm}}
\vskip-12.55mm\hskip8.3mm
\parbox{9cm}{\epsfig{file=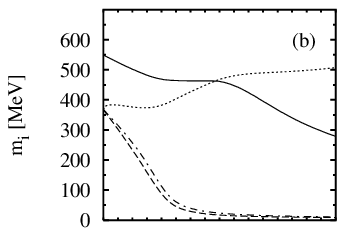, height=6.5cm, width=8.00cm}}
\vskip-13.30mm\hskip8.4mm
\parbox{9cm}{\epsfig{file=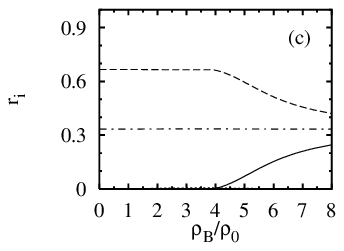, height=7.5cm, width=8.01cm}} \\
\caption{Properties of charge-neutral quark matter with electrons as a 
      function of baryon number density $\rho_B$ (parameters of 
      ref.~\cite{Rehberg}).        
      (a)~Energy per baryon number in chemical equilibrium 
      ($\mu_s = \mu_d = \mu_u +\mu_e$, solid line) and for
      non-strange matter ($\mu_s = 0, \mu_d = \mu_u +\mu_e$, dotted).
      (b)~Dynamical quark masses $m_u$ (dashed-dotted), $m_d$ (dashed) and 
      $m_s$ (solid).
      The dotted line denotes the chemical potential $\mu_s = \mu_d$.
      (c)~$r_i = n_i / n_B$ for the different particle species: 
      up quarks (dashed-dotted), down quarks (dashed), 
      strange quarks (solid), and electrons (dotted).}
\end{center}
\label{fig3}
\end{figure}


\begin{thebibliography}{99}
\bibitem{Bodmer} A.R. Bodmer, Phys. Rev. D 4 (1971) 1601.
\bibitem{Witten} E. Witten, Phys. Rev. D  30 (1984) 272. 
\bibitem{Farhi} E. Farhi and R.L. Jaffe, Phys. Rev. D 30 (1984) 2379.
\bibitem{MIT} A. Chodos, R.L. Jaffe, K. Johnson and C.B. Thorn,
              Phys. Rev. D 10 (1974) 2599;\\
              T. DeGrand, R.L. Jaffe, K. Johnson and J. Kiskis, 
              Phys. Rev. D 12 (1975) 2060.
\bibitem{SQM1} J. Madsen and P.~Haensel (eds.), 
               Strange Quark Matter in Physics and Astrophysics,
               Nucl. Phys. B (Proc. Suppl.) 24B (1991).
\bibitem{SQM2} G. Vassiliadis, A. Panagiotou, B. Shiva Kumar and J. Madsen
               (eds.), Strangeness and Quark Matter
               (World Scientific, Singapore, 1995).
\bibitem{SQM3} Proc. International Symposium on Strangeness
               in Quark Matter (Santorini 1997), J.~Phys. G 23 (1997).
\bibitem{SQM4} C. Greiner and J. Schaffner-Bielich, 
               preprint nucl-th/9801062.  
\bibitem{Buballa} M. Buballa, Nucl. Phys. A 611 (1996) 393;\\
                  M. Buballa and M. Oertel, Nucl. Phys. A 642 (1998) 39c.
\bibitem{NJL} Y. Nambu and G. Jona-Lasinio, Phys. Rev. 122 (1961) 345;
              124 (1961) 246.
\bibitem{HK} T. Hatsuda and T. Kunihiro, Phys. Lett. B 198 (1987) 126.
\bibitem{BJM} V. Bernard, R.L. Jaffe and U.-G. Mei{\ss}ner,
              Nucl. Phys. B 308 (1988) 753.
\bibitem{Taki} M. Takizawa, K. Tsushima, Y. Kohyama and K. Kubodera,  
               Nucl. Phys. A 507 (1990) 611.
\bibitem{Klimt} S. Klimt, M. Lutz, U. Vogl and W. Weise, 
               Nucl. Phys. A 516 (1990) 429; \\
               U. Vogl, M. Lutz, S. Klimt and W. Weise, Nucl. Phys. A 516
               (1990) 469. 
\bibitem{Lutz} M. Lutz, S. Klimt and W. Weise, Nucl. Phys. A 542 (1992) 521.
\bibitem{NJLrev1} U. Vogl and W. Weise, Progr. Part. Nucl. Phys. 27
                  (1991) 195.
\bibitem{NJLrev2} S.P. Klevansky, Rev. Mod. Phys. 64 (1992) 649.
\bibitem{NJLrev3} T. Hatsuda and T. Kunihiro, Phys. Rep. 247 (1994) 221.
\bibitem{Jaminon} M. Jaminon and B. Van den Bossche, 
                  Nucl. Phys. A 582 (1995) 517; \\
                  J. Cugnon, M. Jaminon and B. Van den Bossche,
                  Nucl. Phys. A 598 (1996) 515.
\bibitem{VdB}  B. Van den Bossche, preprint nucl-th/9807010.
\bibitem{Rehberg} P. Rehberg, S.P. Klevansky and J. H\"ufner,
                  Phys. Rev. C 53 (1996) 410.
\bibitem{Barranco} M. Barranco and J.R. Buchler, Phys. Rev. C 22 (1980) 1729.
\bibitem{MS} H. M\"uller and B.D. Serot, Phys. Rev. C 52 (1995) 2072.
\bibitem{Glendenning} N.K. Glendenning, Phys. Rev. D 46 (1992) 1274.
\bibitem{Mueller} H. M\"uller, Nucl. Phys. A 618 (1997) 349.
\bibitem{Chak} S. Chakrabarty, Phys. Rev. D  43 (1991) 627.
\bibitem{Benv} O.G. Benvenuto and G. Lugones, Phys. Rev. D 51 (1995) 1989.
\bibitem{Schertler} K. Schertler, C. Greiner, M.H. Thoma, 
                    Nucl. Phys. A 616 (1997) 659.
\bibitem{ARW}  M. Alford, K. Rajagopal and F. Wilczek, 
               Phys. Lett. B 422 (1998) 247.
\bibitem{Rapp} R. Rapp, T. Sch\"afer, E.V. Shuryak and M. Velkovsky, 
               Phys. Rev. Lett. 81 (1998) 53.
\bibitem{Alford} M. Alford, K. Rajagopal and F. Wilczek, 
                 Nucl. Phys. B 537 (1999) 443.
                 
\end{thebibliography}
\end{document}